\RequirePackage{fix-cm} 
\documentclass[11pt,a4paper]{article}

\usepackage[margin=0.85in]{geometry}
\usepackage{lmodern}
\usepackage[T1]{fontenc}
\usepackage[utf8]{inputenc}
\DeclareUnicodeCharacter{1EF3}{\`y}

\usepackage{graphicx}
\graphicspath{{Figures/}}
\usepackage[labelfont={bf,large},labelsep=period]{caption}
\usepackage[position=bottom]{subfig}  
\usepackage[section]{placeins}
\usepackage{float}
\usepackage{rotating}
\usepackage{tabularx,booktabs}
\usepackage{multirow}
\usepackage{makecell}
\usepackage{array}
\usepackage{dcolumn}
\newcolumntype{.}{D{.}{.}{-1}}

\renewcommand{\arraystretch}{1.5}

\usepackage{amsmath,amssymb,amsthm,amsfonts,mathtools}
\usepackage{bm}
\usepackage{siunitx}
\usepackage{upgreek}

\expandafter\def\expandafter\normalsize\expandafter{%
  \normalsize
  \setlength\abovedisplayskip{4pt}%
  \setlength\belowdisplayskip{4pt}%
  \setlength\abovedisplayshortskip{4pt}%
  \setlength\belowdisplayshortskip{4pt}%
}

\usepackage{xcolor}
\definecolor{darkblue}{rgb}{0.0,0.0,0.66}
\definecolor{Gray}{gray}{0.9}
\definecolor{red2}{RGB}{200,53,58}
\newcommand{\rev}[1]{#1} 

\usepackage[hyperfootnotes=false,bookmarksopen]{hyperref}
\hypersetup{
  pdffitwindow=false,
  pdfstartview={XYZ null null 1.00},
  pdfnewwindow=true,
  colorlinks=true,
  linkcolor=red2,
  citecolor=red2,
  urlcolor=red2
}

\usepackage{enumitem}
\setlist[itemize]{itemsep=0.5pt, topsep=0pt}
\setlist[enumerate]{itemsep=0.5pt, topsep=0pt}
\usepackage{setspace}
\usepackage{textcomp}
\usepackage{colortbl}
\usepackage{pdflscape}
\usepackage{tikz}
\usetikzlibrary{decorations}

\usepackage{ragged2e}
\usepackage[export]{adjustbox}
\usepackage{relsize}
\usepackage{titlesec,titletoc}
\titleformat{\section}{\normalfont\Large\bfseries\color{red2}}{\thesection}{1em}{}
\titleformat{\subsection}{\normalfont\large\bfseries\color{red2}}{\thesubsection}{1em}{}

\usepackage{datetime}
\newdateformat{mydate}{\monthname[\THEMONTH] \THEYEAR}
\usepackage{authblk}
\usepackage[toc,page]{appendix}

\usepackage{csquotes}
\usepackage[
  sortcites=false,
  style=authoryear-icomp,
  backend=biber,
  sorting=nyt,
  hyperref=true,
  maxcitenames=2,
  mincitenames=1,
  uniquelist=false,
  maxbibnames=99
]{biblatex}
\renewbibmacro{in:}{}
\usepackage{endnotes}
\let\footnote=\endnote

\usepackage{indentfirst}          
\definecolor{burgundy}{rgb}{0.5,0.0,0.13}
\makeatletter
\renewcommand\@makefnmark{\hbox{\@textsuperscript{\normalfont\color{burgundy}\@thefnmark}}}
\renewcommand\@makefntext[1]{%
  \parindent 1em%
  \noindent
  \hb@xt@1.8em{\hss\@makefnmark\kern0.2em}#1}
\makeatother

\usepackage{comment}

\begin{document}

\begin{titlepage}
\begin{singlespacing}
\renewcommand\theadalign{bc}
\renewcommand\theadfont{\bfseries}
\renewcommand\theadgape{\Gape[4pt]}
\renewcommand\cellgape{\Gape[4pt]}
\renewcommand\Affilfont{\small}
\renewcommand\Authsep{\hspace{0.7cm}}
\renewcommand{\Authands}{\hspace{0.7cm}}

\author[]{Ilan Strauss, Jangho Yang, and Mariana Mazzucato\thanks{Equal contributors (Ilan Strauss and Jangho Yang). Ilan Strauss is Co-Director at the AI Disclosures Project (Social Science Research Council), and a Senior Fellow at the Institute for Innovation and Public Purpose (IIPP), University College London (UCL), London, UK, WC1E 6BT. Jangho Yang is Assistant Professor of Management Sciences in the Faculty of Engineering, University of Waterloo. Mariana Mazzucato is Professor in the Economics of Innovation and Public Value at University College London, and founding director of Institute for Innovation and Public Purpose (UCL). All errors are our own. Contacts: \url{ilan@aidisclosures.org} (Corresponding Author) and \url{j634yang@uwaterloo.ca}.}}

\date{\small{\mydate \today}  \\[0.2in]
}

\title{``Rich-Get-Richer''? Platform Attention and Earnings Inequality using Patreon Earnings Data}

\maketitle
\thispagestyle{empty}

\vspace{-.5cm}
\begin{abstract}
\doublespacing

Patreon allows content creators to monetize additional content from loyal fans. Because it offers minimal native distribution, Patreon earnings largely reflect creators' popularity on their primary external platforms (Instagram, Twitch, YouTube, Twitter/X, Facebook, or ``Patreon-only''), making them a useful proxy for platform-level algorithmic attention dynamics. Fitting power-law tails to Patreon earnings by primary platform affiliation, we find three key results. First, platforms exhibit ``rich-get-richer'' earnings dynamics \parencite{barabasi1999emergence}, reflected in a Pareto exponent $\alpha \approx 2$, which is closer to concentrated capital income than labor income. Second, platforms with more concentrated earnings (lower $\alpha$) have lower mean and median earnings, and thus an eroded creator ``middle class.'' Third, across most platforms, $\alpha$ values decline and converge over time \rev{(based on three cross-sections: 2018, 2021, and 2024)}, meaning earnings become increasingly concentrated among top creators -- consistent with algorithmic recommendations rising in importance. Algorithmic attention allocation is a plausible driver of these patterns, though platform-specific conversion rates and audience willingness to pay may be just as important. 

\end{abstract} \medskip


\end{singlespacing}
\end{titlepage}

\newpage
\pagenumbering{arabic}

\newpage
\section{Introduction}
\doublespacing 
\frenchspacing

Online social media markets are governed by algorithmic allocations of attention \parencite{o2024algorithmic, strauss2024Amazon}, not prices. Unlike classical multi-sided market models \parencite{rochet2003platform, rochet2006two}, where prices balance supply and demand, online platforms rely predominantly on non-price algorithmic mechanisms that process vast data on content and user preferences to generate matches and allocate attention \parencite{cunningham2024we}.

Algorithms, however, don't just create new markets and enable efficiencies \parencite{varian2010computer}, they are distributional devices that decide who benefits and who does not \parencite{hitzig2019technological}. We examine earnings inequality \emph{within} a platform market, where algorithmic recommendations allocate visibility and monetization potential among comparable participants. Platforms that recommend the same `viral' content to all users may generate highly unequal earnings, whereas algorithms that surface niche 'long tail' content may foster more equitable distributions \parencite{anderson2012long, brennan2020attention}. This within-platform distributional channel is distinct from the forces that dominate the broader technology-and-inequality literature, that operate between-workers and between-firms  \parencite{autor2020fall, acemoglu2022tasks}. We, therefore, are providing empirical evidence on just one piece of this larger distributional dynamic. 
  

Unfortunately, there is little detailed evidence of the economic impact of algorithmic attention mechanisms, since platforms do not disclose creator earnings or algorithmic optimizations \parencite{gillespie2018custodians, mazzucato2023regulating, oreilly2023regulating, Naceva_2025_CreatorEP}.\footnote{Social media platforms are not covered by national statistical classifications, such as the U.S. Census Bureau.} Patreon is perhaps the only large creator platform to disclose earnings, but this has not been analyzed systematically to test for cross-platform attention and earnings inequality \parencite{regner2021crowdfunding, hutson2021youtube, el2022quantifying, anderson2022patreon}.

This paper fills this gap by using monthly Patreon earnings to examine attention dynamics generated by third-party social media platform algorithms, revealing whether these algorithms foster unequal ecosystems or support a healthy creator middle class. We exploit a novel feature in a large multi-year Patreon dataset ($n = 104,719$ creators across 2018, 2021, and 2024) that identifies creators who use only a single social media platform (or only Patreon). As a monetization tool for selling additional content to loyal subscribers, Patreon offers little native distribution, making its earnings an ideal proxy for attention captured on external platforms (Instagram, Twitch, YouTube, Twitter/X, Facebook).

With cross-sectional data, we can measure and compare the \emph{shape} of Patreon monetization outcomes across platform-affiliated creators. What we cannot separately identify is algorithmic attention allocation from creator sorting, audience composition, or platform-specific conversion norms. We interpret cross-platform differences as documenting variation in earnings inequality among Patreon creators affiliated with each platform, treating algorithmic attention allocation as one plausible mechanism among several rather than a definitively identified causal channel.

We characterize earnings using power law distributions, $p(x) \sim \frac{1}{x^\alpha}$, which describe systems where small occurrences are extremely common while rare large occurrences have significant impact. A lower power law exponent implies higher inequality, indicating greater probability of extreme tail incomes characteristic of wealth distributions \parencite{pareto1896cours, atkinson1970distribution}. Such heavy-tail distributions also occur in internet traffic, where a few websites capture most visits \parencite{adamic2002zipf, huberman1998internet}.

\rev{Our core finding is that monthly Patreon earnings across platforms fit a power law with exponent around $\alpha \approx 2.0$, indicating an extremely heavy-tailed distribution and a strong ``rich-get-richer'' dynamic. This exponent is consistent with the copying model of preferential attachment \parencite{mitzenmacher2004brief, easley2010networks, chung2003generalizations} operating near its pure-preferential limit.}\footnote{Rather than the original Barab\'asi-Albert model \parencite{barabasi1999emergence}, which predicts $\alpha = 3$.} This places creator earnings closer to concentrated capital income ($\alpha \approx 1.5$) \parencite{gabaix2009power, kumar2024power} than to labor income, suggesting multiplicative and compounding effects drive online creator earnings \parencite{strauss2024Amazon}.\footnote{For debate see: \cite{brzezinski2014wealth}.} Platform variation is consistent with differences in algorithmic systems: YouTube ($\alpha = 1.8$), Instagram ($\alpha = 1.84$), Twitch ($\alpha = 1.93$), and Facebook ($\alpha = 1.94$). 

Second, platforms with stronger power laws (lower $\alpha$), such as YouTube and Instagram, exhibit weaker creator middle classes: lower mean and median earnings with less expansive interquartile ranges (Q25-Q75). The ``rich-get-richer'' dynamic draws gains disproportionately from mid-tier creators. In contrast, platforms with weaker power laws—Patreon and especially Twitter—support healthier middle-class creator ecosystems, which may reflect algorithmic differences in content discovery, though audience composition and platform-specific monetization norms could also contribute.

Third, \rev{comparing three cross-sections (2018, 2021, 2024),} we observe convergence toward steeper inequality across YouTube, Twitch, Instagram, and Facebook, with declining $\alpha$ values implying increasing earnings concentration. Instagram shows the most extreme shift, with $\alpha$ worsening from around 2.1 in 2021 to 1.84 in 2024, which is consistent with a growing role for algorithmic recommendations relative to social-graph mechanisms \parencite{mignano2022recommendation}, though compositional changes in the creator population offer an alternative explanation.

How should we interpret these findings given that Patreon earnings represent only one revenue stream? Our data provides a partial view of total creator earnings inequality and so should not be interpreted as representative of total earnings. That being said, the major creator revenue sources -- sponsorships, affiliate marketing, ads -- likely exhibit similar (or greater) inequality since they depend on the same attention mechanism: more popular creators earn more \parencite{imh_creator_earnings_2023, rieder2023making, oreilly2023regulating}.\footnote{Patreon monetization likely represents a small component overall. One study finds 2.55\% of YouTube video descriptions contain crowdfunding links versus 49\% for cross-platform links.} Our earnings data most closely proxies for algorithmic attention mechanisms on third-party platforms where these creators are active, since Patreon monetization relies almost entirely on external platforms' algorithms to generate attention, as Patreon offers minimal native discovery capabilities.\footnote{Patreon's creator search functionality is primitive.} Supporting this interpretation is our finding that creators with no other social media accounts (Patreon-only) show the second least concentrated distribution (highest $\alpha$), after Twitter/X -- whose historical reliance on social graphs rather than pure recommendation may explain its relatively equitable outcomes.

A further caveat concerns the conversion rate from platform attention to Patreon subscriptions, which may vary across platforms for reasons unrelated to algorithmic attention allocation. We discuss this confound formally in Section~\ref{sec:conversion_confound}.

One interpretation of declining $\alpha$ values is that platforms' algorithmic allocations have become more concentrated, which may be associated with increasing market power \parencite{strauss2024Amazon, o2024algorithmic, Doctorow2024} -- though the relationship between within-platform earnings allocations (driven by algorithmic attention) and market structure is not straightforward (Section \ref{sec:conclusion}). Potential corrective measures (Section \ref{sec:conclusion}), such as mandating recommendation systems that better incentivize long-tail content creation \parencite{park2008long, hu2023incentivizing, yao2024unveiling}, may effectively foster more equitable creator ecosystems. And to the extent that unequal earnings distributions reflect a platform's algorithmic market power \parencite{o2024algorithmic}, antitrust policy may have a role to play too.\footnote{But if algorithmic addiction suspends rational consumer choice then consumer protection laws may be a more appropriate algorithmic governance channel.}

Despite previous studies employing Patreon data \parencite{regner2021crowdfunding, el2022quantifying, hutson2021youtube}, we provide the first systematic analysis of the earnings distribution's underlying structure and the economic processes or algorithmic mechanisms generating it. We also provide some of the only evidence on content creator earnings distributions \parencite{influencermarketinghub2023creator, peres2024creator, Naceva_2025_CreatorEP}.


Section~\ref{sec:data_method} presents our data and power law methodology, including the ``rich-get-richer'' dynamic underpinning it (discussed further in Appendices). Section~\ref{sec:results} presents our findings. Section~\ref{sec:Discussion} discusses policy implications: how to incentivize greater incorporation of long-tail content \parencite{anderson2012long} in recommendation systems, and the impact of market power on this. Section~\ref{sec:conclusion} concludes. Robustness and data imputation are discussed in Appendix~\ref{app:robust}.

\section{Data and Method}
\label{sec:data_method}

This section motivates our use of a power law to estimate earnings, and our interpretation of it as reflecting a `rich-get-richer' dynamic. \rev{We draw on the copying model of preferential attachment \parencite{mitzenmacher2004brief, easley2010networks, chung2003generalizations}, which predicts that attention follows a power law distribution with density \( p(A) \propto A^{-\alpha} \) and can generate exponents around $\alpha = 2$ when attention is allocated almost entirely by existing popularity.} Before discussing the model we first detail our data.

\subsection{Data and Descriptive Statistics}

We use official Patreon monthly earnings data for content creators purchased directly from the Patreon platform, for the month of March for years 2018, 2021, and 2024 (n = 104,719).

Patreon is a monetization platform for content creators, allowing them to sell additional or primary content to loyal subscribers (paid and unpaid). This is done by linking to their Patreon from their existing social media profiles. Viewers discover these links when they encounter the creator's content or visit their profile on the third-party social media site, such as Instagram or YouTube. Patreon typically appeals to creators with close follower relationships. Other common link-driven monetization strategies by creators on platforms include external links to Amazon, Spotify, Etsy, or App Stores, cross-promotion of the creator's other content on the same platform, or using internal platform monetization tools \parencite{rieder2023making}. 

For most creators, Patreon is therefore only one component of total earnings; creators also earn directly from their primary platform through ads, affiliate programs, and sponsorships (creators whose only affiliation is Patreon are the exception). These other revenue streams are likely at least as unequally distributed as Patreon earnings -- if not more so -- since they appear to scale exponentially with viewership \parencite{influencermarketinghub2023creator}. Unfortunately, no detailed public data on these earnings exist.

Patreon, however, allows creators to keep their earnings private, leading to significant missing data here, especially for larger earners, since those with more paid subscribers tend to disclose their earnings the least, we find. Not imputing missing earnings, therefore, would introduce a significant downward bias on the right tail of the distribution. We impute missing earnings linearly using a basic specification, based on earnings being a function of paid members (subscribers), all members, the type of content created, if they are an NSFW content creator, and the year of the data. We show in Appendix~\ref{sec:robust} that the method is robust and has good predictive power, despite its simplicity. We then restrict the dataset to include only content creators with monthly earnings exceeding \$10 USD. 

Next, we segment the data by the social media platform(s) the Patreon content creators use to engage their audiences. We identify creators who only use a single social media platform and remove from our dataset all earners with more than one social media affiliation. The dataset spans Patreon creators across major platforms, including YouTube, Instagram, Twitter/X, Twitch, Facebook, and Patreon only. Table~\ref{tab:summary} provides summary statistics for earnings by platform, detailing the number of observations, mean earnings, standard deviation (SD), and key percentiles (minimum, 25th, 50th, 75th, and maximum).\\[-2mm]

\begin{table}[H]
\begin{center}
\caption{\centering {\large{Summary Statistics: Patreon monthly earnings by social media platform}}} 
\vspace{-2mm}
\label{tab:summary}

\begin{tabular}{llllllllll}
  \toprule
Platform & Obs & Mean & Median & SD & Min & Q25 & Q50 & Q75 & Max \\ 
  \midrule
Facebook & 8,458 & \$149 & \$47 & \$631 & \$10 & \$30 & \$47 & \$99 & \$35,261 \\ 
  Instagram & 35,879 & \$226 & \$59 & \$1,076 & \$10 & \$33 & \$59 & \$142 & \$60,391 \\ 
  Patreon & 83,950 & \$278 & \$57 & \$1,667 & \$10 & \$31 & \$57 & \$154 & \$211,321 \\ 
  Twitch & 1,793 & \$198 & \$46 & \$1,174 & \$10 & \$29 & \$46 & \$105 & \$31,436 \\ 
  Twitter & 47,564 & \$373 & \$72 & \$2,159 & \$10 & \$34 & \$72 & \$215 & \$169,106 \\ 
  Youtube & 23,435 & \$250 & \$47 & \$1,490 & \$10 & \$28 & \$47 & \$119 & \$87,802 \\ 
   \bottomrule
\end{tabular}


\end{center}
\vspace{-6mm}
\caption*{\textit{Note}: The table includes the number of observations (Obs), mean earnings, standard deviation (SD), and percentiles (Min, Q25, Q50, Q75, Max) for creators on each platform in monthly US\$.}
\end{table}

The data show substantial variation in earnings across platforms. Twitter/X and Patreon have the highest mean, median, maximum, and interquartile range of earnings in our sample, while Facebook and Twitch sit at the lower end. At least part of this pattern, especially the very high maxima, is likely driven by the much larger number of creators we observe on Twitter/X and Patreon, rather than by systematically different underlying inequality across platforms.\footnote{Raw maxima and interquartile ranges can both be misleading indicators of inequality when outcomes have very heavy upper tails, as is the case for creator earnings. Intuitively, a platform with more creators has many more ``draws from the lottery'' of extreme success. Even if two platforms had identical underlying earnings distributions, the one with more creators would tend to exhibit a higher observed maximum simply because it is more likely to contain an extremely successful creator. In the following sections we show that creator earnings are well described by a power-law (Pareto-type) distribution. In such distributions, the expected maximum in a sample of size $N$ grows quickly with $N$ (roughly like $N^{1/(\alpha-1)}$ in a Pareto model), so larger platforms naturally produce more extreme top observations even when the underlying tail parameter $\alpha$, our measure of inequality, is the same. At the same time, the interquartile range focuses on the middle of the distribution and ignores the very top tail, so platforms with very different degrees of top-end concentration can have similar interquartile ranges. This is why, for characterizing inequality, we focus on the estimated exponent $\alpha$ rather than on raw maxima, interquartile ranges, or other platform-specific summary statistics.}  We note later that this proxy for ecosystem health or strength (median, mean, and interquartile earnings), correlates strongly with the estimated power law for each platform. The standard deviation of earnings is highest for Twitter/X and lowest for Facebook. Facebook is the only platform to have fewer exclusive Patreon content creators in our data in 2024 than it had in 2021 (Table \ref{tab:summary2} in Appendix).

\subsection{Platform-to-Patreon conversion as a confound}
\label{sec:conversion_confound}

A key interpretive challenge in using Patreon earnings to study cross-platform attention dynamics is that we observe monetization outcomes on Patreon, not attention directly on the originating platform. For a creator $i$ affiliated with platform $p$, monthly Patreon earnings $Y_{ip}$ can be thought of as the product of three components:
(i) algorithmic attention allocations $A_{ip}$ that reinforce advantage or tap into a longer-tail of creators and content; and then two potential confounding variables: (ii) a platform-to-Patreon conversion component $C_{ip}$ covering the propensity for a unit of platform attention to translate into a paid patron action, and (iii) an average revenue-per-conversion component $R_{ip}$ that could be shaped by different users being on each platform. A simple decomposition of this is:
\[
Y_{ip} \;=\; A_{ip}\times C_{ip}\times R_{ip}.
\]
Or: $\log Y_{ip}=\log A_{ip}+\log C_{ip}+\log R_{ip}$. Our data directly identify the distribution of $Y_{ip}$, but do not separately identify the distributions of $A_{ip}$, $C_{ip}$, and $R_{ip}$.

$C_{ip}$ and $R_{ip}$ might vary systematically across platforms for reasons unrelated to algorithmic attention allocations, $A_{ip}$. Conversion can differ due to audience norms (``support culture''), the friction of linking out, the prominence of links in profiles, the share of creators using Patreon as a primary versus supplemental income stream, and the availability of substitute monetization (e.g., ad revenue sharing, tipping, subscriptions, or brand deals). These differences can mechanically generate cross-platform variation in observed Patreon earnings inequality even if underlying attention distributions were identical.

The extent to which this confound affects our core inequality statistic (the estimated power-law exponent $\alpha$) depends on how conversion varies with creator scale. If platform-level conversion and revenue-per-conversion operate approximately as a multiplicative scale factor, then our interpretation focusing on $A_{ip}$ is sensible. For example, if $C_{ip}R_{ip}\approx s_p$ is roughly constant across creators within platform $p$, then $Y_{ip}\approx s_p A_{ip}$ and cross-platform comparisons of tail exponents are more interpretable; since multiplying by a constant shifts the earnings level but does not change the tail shape. By contrast, if conversion is systematically \emph{rank-dependent}, meaning, if larger creators convert platform attention into patrons at higher rates, because they cultivate stronger parasocial relationships, have better funnel design, or face different audience composition), then $C_{ip}$ may increase with $A_{ip}$ and can steepen or flatten the observed tail. For instance, if $C_{ip}R_{ip}$ scales with attention as $(C_{ip}R_{ip})\propto A_{ip}^{\beta}$ within a platform, then $Y_{ip}\propto A_{ip}^{1+\beta}$ and the observed earnings tail can differ from the underlying attention tail even when attention allocation mechanisms are unchanged.

For these reasons, we interpret cross-platform differences in $\alpha$ conservatively as differences in the \emph{inequality of Patreon monetization outcomes among creators affiliated with each platform}. These patterns are consistent with more or less concentrated algorithmic attention allocations by each platform, but they could also reflect platform-specific conversion regimes or creator sorting into Patreon as a monetization channel.\footnote{In the robustness analysis, we therefore emphasize diagnostics that speak to whether the platform ordering of $\alpha$ is driven primarily by earnings levels (a scale shift) versus changes in tail shape, and we treat claims about algorithmic attention allocation as suggestive rather than definitive in the absence of direct attention measures or quasi-experimental variation in platform policies.}

\rev{Several features of the data nonetheless support the plausibility of cross-platform comparisons. First, all Patreon creators use the same monetization mechanism -- a subscription paywall -- holding the conversion \emph{tool} constant even if the propensity to convert varies across audiences. Second, the temporal trends we document within platforms (declining $\alpha$ for Instagram, YouTube, Facebook, and Twitch) cannot easily be explained by fixed cross-platform differences in audience culture, since those would produce level differences rather than within-platform trends. Third, content categories on Patreon are broadly similar across platform-affiliated creators (see Table~\ref{tab:summary}), suggesting creator populations are not so compositionally different as to invalidate comparison. Fourth, if conversion culture alone drove the observed patterns, we might expect platforms with larger creator bases to show more diverse earnings distributions through sheer variety; yet YouTube-affiliated creators show the \emph{most} concentrated earnings despite YouTube's dominance as a traffic source, suggesting that attention allocation mechanisms are a plausible driver alongside conversion norms.}

\subsection{Power Law Model}
\label{sec:datanmethod}
Power law distributions, $p(x) \sim \frac{1}{x^\alpha}$, provide a useful framework for characterizing earnings online, offering insights into system scaling, earnings inequality, and underlying mechanisms. These distributions describe systems where small occurrences are extremely common while large occurrences are extremely rare but have significant impact. A lower power law exponent implies higher inequality, indicating greater probability of extreme tail incomes characteristic of long-tail distributions such as wealth \parencite{pareto1896cours, atkinson1970distribution} or internet traffic, where a few websites capture most visits \parencite{adamic2002zipf, huberman1998internet}.

Much evidence shows that various aspects of society and the economy follow power laws \parencite{gabaix2009power}, with the vast majority of outcomes small yet a few extraordinarily large, commanding substantial shares of the total. Online markets often exhibit similar extreme disparities, with a small number of top content creators or e-commerce sellers capturing disproportionate revenue shares \parencite{fortunato2006topical, clauset2009power}.

The hypothesis we estimate is that ``rich-get-richer'' dynamics, characterized by preferential attachment models \parencite{barabasi1999emergence}, govern content creator earnings. We emphasize preferential attachment because it captures the global algorithmic mechanism: recommendation systems explicitly allocate attention in proportion to existing popularity. Related copying models, where agents imitate others' connections locally \parencite{kleinberg1999web, kumar2000stochastic}, can produce similar power-law outcomes and may interact with algorithmic amplification in practice.

As creators gain visibility, they attract further attention through algorithmic amplification \parencite{zappin2022youtube, hua2022characterizing}, creating a self-reinforcing feedback loop that concentrates income among a small group of highly popular creators -- a ``winner takes all'' or viral effect \parencite{rosen1981economics, rieder2018ranking}. This compounding of initial advantage, also called a ``superstar effect'' in economics \parencite{gabaix2009power, autor2020fall}, means that participants who accumulate more engagement become even more likely to attract future interactions. Patreon earnings data provide concrete, measurable outcomes of these dynamics given their reliance on third-party platform algorithms for visibility.

A similar dynamic of persistence governs firm dynamics in the broader economy \parencite{geroski2001modelling, geroski2002learning, bottazzi2006explaining}; however, online, network and algorithmic effects can supercharge this \parencite{easley2010networks}.\footnote{The low marginal cost to distributing and selling content online adds to the power law, which arises from the feedback introduced by correlated decisions across a population \parencite{easley2010networks}.}\footnote{Some econophysics models distinguish a two-class structure in income distributions, where working class income follows an exponential distribution (reflecting additive processes) and rentier or capitalist class income follows a power law (reflecting multiplicative processes). See \parencite{ludwig2022physics, druagulescu2001exponential}.}

Detecting power law distributions in earnings data provides valuable insights into the health and fairness of online ecosystems. Highly skewed distributions, with large numbers earning very little, may signal systemic challenges such as algorithmic bias, barriers to entry, or lack of upward mobility \parencite{anderson2022patreon, needleman2023influencers}. Conversely, distributions less dominated by top earners suggest more inclusive and competitive environments with broader opportunities for success. Understanding these patterns can inform platform governance, guide policymakers, and help researchers propose fairer algorithms and revenue-sharing mechanisms, ultimately leading to more equitable digital economies \parencite{eckhardt2023creator, gillespie2018custodians}.

\subsection{Power Laws and Algorithmically-Driven `Rich-Get-Richer' Dynamics}

A power law can be written as:
\begin{equation}
p(A) \propto A^{-\alpha},
\end{equation}

where \( P(A) \) is the probability of observing a node (e.g., a content creator) with attention \( A \) (such as clicks, views, or follows), and the exponent \(\alpha\) characterizes the degree of inequality in how attention is distributed across the network.

Power law distributions appear in many domains \parencite{gabaix2009power}, often driven by self-reinforcing processes. One such process is captured by the model of \textcite{barabasi1999emergence}, which illustrates how compounding algorithmic advantage emerges in social and digital networks. The foundational mechanism resembles a reinforced P\'olya urn model \parencite{chung2003generalizations, chung2006concentration}, where the probability of adding a ball of a certain color increases with the number already in the urn. In the digital realm, each new click or view is more likely to go to creators who already have substantial visibility, resulting in a `rich-get-richer' dynamic commonly observed on platforms governed by algorithmic curation \parencite{barabasi2005origin, jackson2008social, gabaix2009power, vidal2011interactome, easley2010networks, barabasi2011network, barbosa2018human}.

To model this, we introduce a parameter \(\gamma\) that governs the balance between random algorithmic attention allocation (exploration) and attention guided by existing popularity (exploitation):

\begin{itemize}
    \item With probability \(\gamma\), an attention event (e.g., a new user's click, view, or follow) is directed to a creator chosen uniformly at random.
    \item With probability \(1 - \gamma\), attention is directed proportionally to a creator's existing attention level.
\end{itemize}

This parameter captures how platform algorithms trade off exploration (surfacing less-known creators) against exploitation (reinforcing established winners). A low \(\gamma\) implies strong preferential attachment to already-popular creators, whereas a high \(\gamma\) distributes attention more evenly and weakens the rich-get-richer effect.

In the original Barabási–Albert (BA) model there is no random component: new links attach purely in proportion to existing degree, which generates a power-law degree distribution with
density exponent \(\alpha = 3\).\footnote{ In this mixed model, choosing \(\gamma = 1/2\) yields the same exponent \(\alpha = 3\), although the underlying mechanism differs from the original BA construction. See Appendix~\ref{app:ba_model} for the standard BA derivation.} A closely related BA-style mixed (copying) model, in which attention is allocated according to the two rules above, yields a power-law density \(p(A) \propto A^{-\alpha}\) with\footnote{See Appendix~\ref{app:power_law_der} for more details.}
\begin{equation}
\alpha = 1 + \frac{1}{1 - \gamma}.
\label{eq:alpha_gamma}
\end{equation}

Within this mixed model, reducing \(\gamma\) (relying more on popularity-driven exploitation) lowers \(\alpha\), intensifying inequality and concentrating attention among top creators, while increasing \(\gamma\) raises \(\alpha\) and produces a more balanced distribution of attention.

\subsection{Interpretation of \(\alpha\)}

The exponent \(\alpha\) provides insights into algorithmically mediated attention ecosystems. Since \(\alpha\) is directly shaped by \(\gamma\), it reflects how platform algorithmic preferences influence overall attention distribution:

\begin{itemize}
    \item A smaller \(\alpha\) (closer to 2) indicates algorithms heavily favor already popular creators (lower \(\gamma\)), leading to steeply skewed attention distributions.
    \item A larger \(\alpha\) corresponds to more equitable attention distribution, arising from higher \(\gamma\) algorithms that are more exploratory and less biased toward existing popularity.
\end{itemize}

Applied to platform earnings, the same attention-reinforcing mechanisms shape creators' revenue streams. If visibility and engagement are governed by these dynamics, then earnings -- including earnings derived from ads, sponsorships, or subscriptions -- also become concentrated among a few popular creators. By estimating \(\alpha\) from observed earnings distributions, we can quantify how deeply entrenched the rich-get-richer effect is in these digital ecosystems and assess whether algorithmic changes (modifying \(\gamma\)) can create more sustainable conditions for a broader ``middle class'' of creators.

\rev{An important caveat is that \(\gamma\) should be understood as a reduced-form parameter capturing the \emph{net effective} exploration rate, not a direct measurement of any specific platform mechanism. Real recommendation systems inject exploration through cold-start boosts, diversity penalties, and randomized A/B slots. What \(\alpha \approx 2\) tells us is that the net effect of all algorithmic interventions, user behavior, and feedback loops produces an outcome consistent with very little effective exploration, even if individual components of the system are more exploratory than \(\gamma \to 0\) would suggest.}

\section{Results}
\label{sec:results}
To fit the power law tail, we apply the method introduced by \textcite{clauset2009power}, which relies on maximum likelihood estimation to fit the power law model and employs the Kolmogorov-Smirnov (KS) statistic to assess the goodness-of-fit. This approach estimates the scaling parameter by maximizing the likelihood of the observed data and determines the lower bound for power law behavior by minimizing the distance between the empirical and theoretical distributions. The analysis is conducted using the \texttt{poweRlaw} package \parencite{poweRlaw} in R \parencite{R_core_team_2024}.\\[-2mm] 

\begin{figure}[H]
	\begin{center}
    \caption{\centering {\large{Log-log complementary cumulative distribution function (CCDF) of earnings data for each platform with fitted power law lines.}}}
	\begin{minipage}{1.0\textwidth}
		\includegraphics[width=\textwidth]{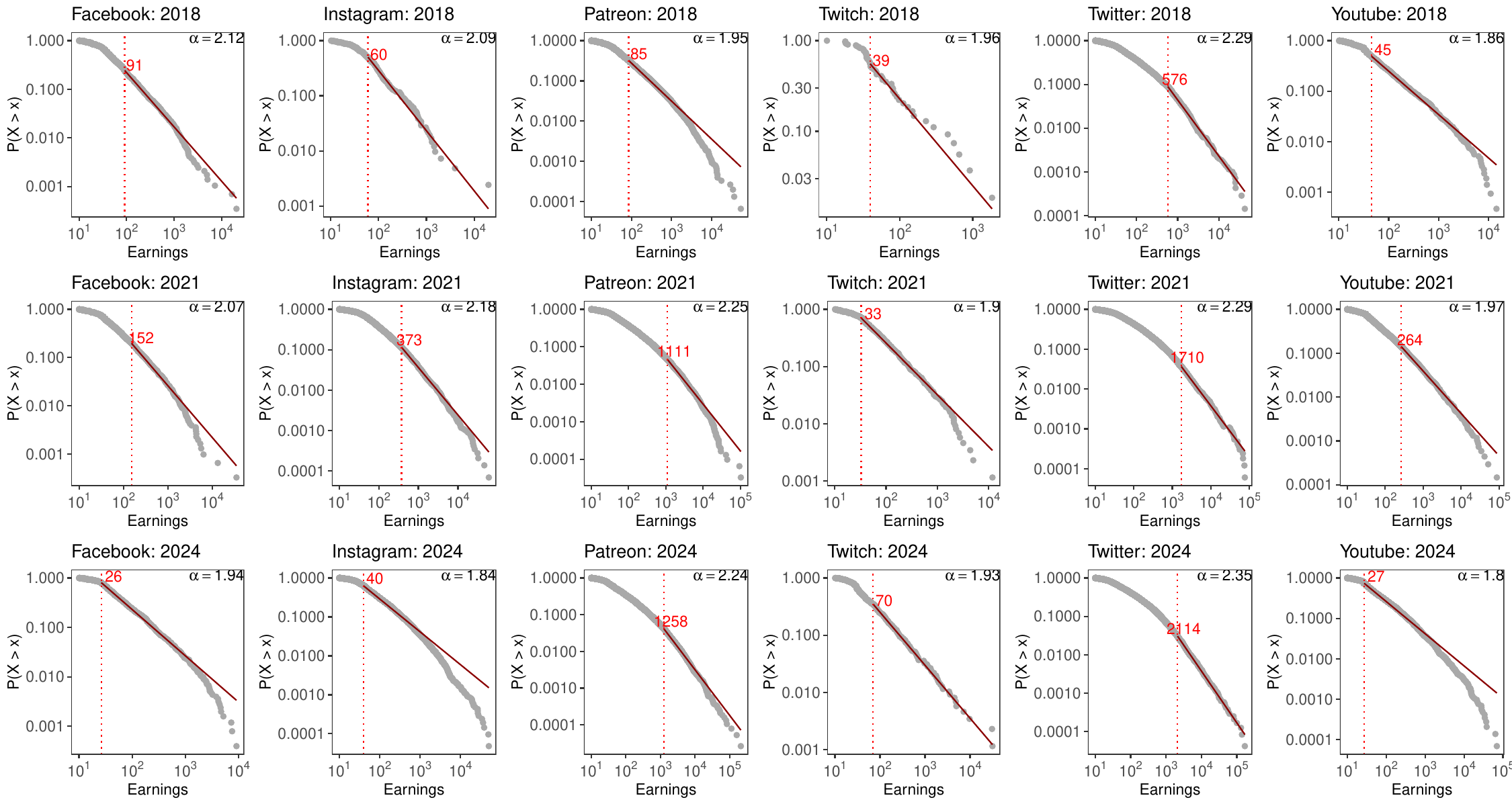}
	\end{minipage}
    \vspace{2mm}
  \caption*{\textit{Note}: The red vertical dotted line indicates the estimated minimum value at which the distribution follows a power law pattern. Estimations were performed for each platform and year, and results remained robust across resampled datasets.}
	\label{fig:power_law_est}
	\end{center}
\end{figure}

Figure~\ref{fig:power_law_est} displays the log-log complementary cumulative distribution function (CCDF) of earnings data across the major social media platforms YouTube, Instagram, Twitch, Patreon, Twitter/X, and Facebook, along with the ideal power law lines for each in red (sloping downwards). The red vertical dotted line represents the estimated minimum value above which the earnings distribution adheres to a power law pattern. Several key findings emerge.\footnote{See Appendix~\ref{app:robust} for a discussion on the robustness of the results.}

The earnings distributions for most platforms in Figure~\ref{fig:power_law_est} exhibit a strong power law pattern around $\alpha=2$.\footnote{
Formally, if individual earnings $X$ follow a Pareto-type distribution
with density $p(x) \propto x^{-\alpha}$ for $x \ge x_{\min}$ and
$\alpha > 1$, then the upper tail behaves as
$P(X \ge x) \propto x^{-(\alpha - 1)}$. Thus, for $\alpha \approx 2$,
the probability of observing earnings above a level $x$ falls roughly
inversely with $x$: increasing earnings by a factor of 10 makes such
high earners about 10 times rarer. For example, if 10{,}000 creators earn at least \$100, we would expect around 1{,}000 to earn at least \$1{,}000, about 100 to earn at least \$10{,}000, and roughly 1 to earn at least \$1{,}000{,}000. This follows from the scale invariance properties of the power law coupled with our finding that the power law begins very early on in the earnings distribution. This means that the benefits to moving up the algorithmic `ladder' are likely enormous at each rung, highlighting just how important it is to be higher up the screen, or the user's attention sphere. Position matters enormously to monetization.} YouTube, Instagram, Twitch, and Facebook consistently show well-defined power law distributions, ranging between 1.8 and 1.94 for the year 2024, suggesting significant earnings concentration among top creators. Patreon and Twitter/X show a much weaker power law (2.24 and 2.35, respectively). 

Power law dynamics emerge early at between \$26 and \$70 per monthly earnings across these platforms, a very low monthly earnings amount consistent across platforms. At an annual amount of \$312 to \$840, this is in stark contrast to standard income distributions, where the power law typically applies to income levels only above \$100,000 per annum \parencite{druagulescu2001exponential, tao2019exponential}. 

\begin{figure}[H]
	\begin{center}
      \caption{\centering {\large{Average $\alpha$ by platform (pooled years) and $\alpha$ time series for each platform.}}}
		\begin{minipage}{.99\textwidth}
		\includegraphics[width=\textwidth]{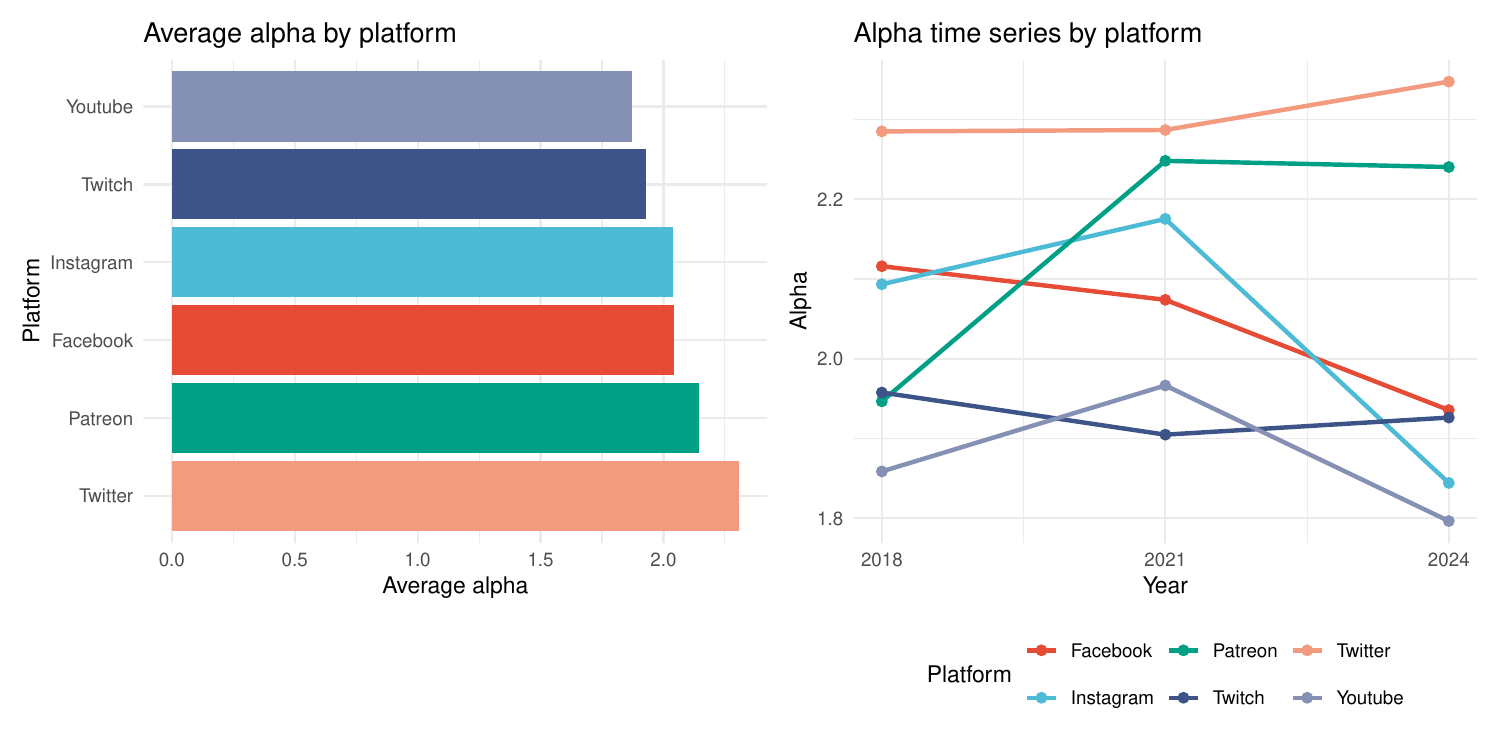}
	\end{minipage}

	\label{fig:alpha_analysis}
	\end{center}
    \vspace{-8mm}
  \end{figure}

Unlike income distribution models where the power law typically governs only the extreme right tail \parencite{tao2019exponential},\footnote{With the majority of the distribution better explained by exponential or gamma-type distributions.} our results show that social media earnings data are largely dominated by a power law from very early on in the earning distribution. It is worth noting though that the extreme right tail is not as concentrated as a pure power law for most platforms, represented by the grey data points in Figure~\ref{fig:power_law_est} falling off below the slanted red power law lines. Twitter/X, however, shows a strong power law on its right-most tail, implying that earnings become extremely concentrated but only once a higher threshold is reached (\$2,114). This means that opportunities for Patreon content creators on Twitter/X are likely to be greater than other comparable platforms below this threshold (especially for `NSFW' creators). Figure~\ref{fig:alpha_analysis} plots the average $\alpha$ by platform (left hand side) and year (right hand side).\\[-2mm] 

Platform-specific power law exponents ($\alpha$) exist and are consistent with differences in earnings inequality across platform-affiliated creators, though they may also reflect audience composition, creator sorting, or conversion norms. YouTube and Twitch have the lowest, i.e. most extreme, estimated $\alpha$ values, suggesting relatively heavier tails and higher inequality, indicating a greater tendency for the emergence of super-hubs—nodes with degrees larger than expected in the standard preferential attachment model \parencite{barabasi1999emergence}. Twitter/X and Patreon consistently display the highest (least extreme) $\alpha$ values, reflecting lighter tails and less earnings inequality.

Furthermore, temporal trends in $\alpha$ show a worsening in the platform power-law relation, which is consistent with (though not definitive evidence for) growing concentration in algorithmic attention allocation. Across Facebook, Instagram, Twitch, and YouTube, $\alpha$ values have generally worsened over time (getting smaller), reflecting a transition toward heavier tails and higher earnings inequality at the extremes, as a pure Pólya urn effect emerges in line with the preferential attachment model \parencite{barabasi1999emergence}. YouTube has the lowest value in 2024, followed closely by Instagram, with values around 1.8 and 1.88. Lastly, Twitter/X's $\alpha$ increased (indicating less concentration) between 2021 and 2024 -- a period spanning Elon Musk's acquisition of the platform in October 2022. We do not attempt to attribute this pattern to any specific platform change; it may reflect selection and composition shifts (including creator migration and the influx of NSFW creators) as much as algorithmic or policy differences. Though `NSFW' (not safe for work) creators increased on the platform by almost 30 percentage points, going from 45\% and 46\% of Patreon creators on Twitter/X in 2018 and 2021, respectively, to 58\% in 2024, reflecting a growing permissiveness on the platform (Table \ref{tab:summary2} in Appendix).\footnote{Note that these time-series patterns should be interpreted cautiously: the sample consists of only three cross-sections (2018, 2021, 2024), and a few platform-year estimates, especially Patreon and Instagram in the final year, are somewhat sensitive to the estimation method used (see Appendix~\ref{app:robust}).}\\[-2mm]

\begin{figure}[H]
	\begin{center}
      \caption{\centering {\large{Median vs. $\alpha$ across platforms.}}}
	\begin{minipage}{.69\textwidth}
		\includegraphics[width=\textwidth]{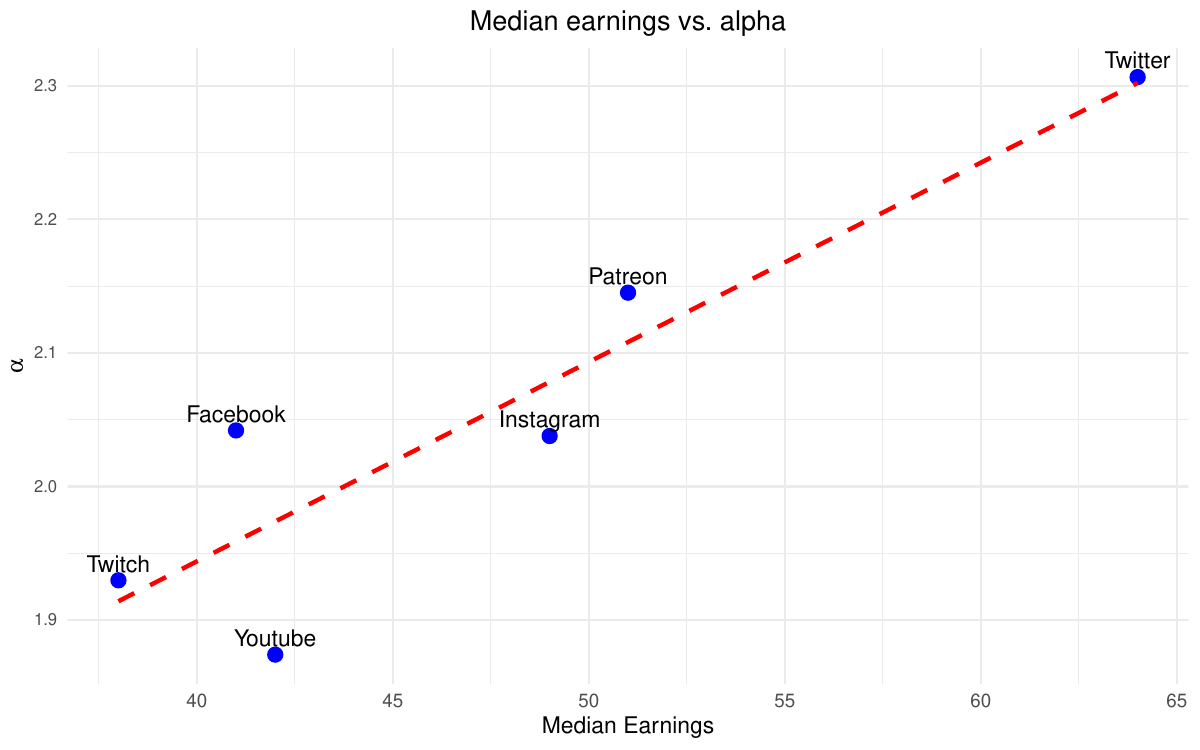}
	\end{minipage}

	\label{fig:scatter_median}
	\end{center}
         \vspace{-4mm}
  \end{figure}

Figure~\ref{fig:scatter_median} shows a clear association between the power law exponent ($\alpha$) and the median of the earnings distribution: platforms with higher median earnings tend to exhibit less heavy-tailed distributions (higher $\alpha$). This indicates that high earnings for the middle layer of creators (``middle-class'') are associated with less unequal platforms. Twitter/X -- and to some extent Patreon -- stands out in this context, combining a high median income with a relatively equal earnings distribution up until a certain point. Several explanations may contribute to this pattern beyond algorithmic design. Twitter/X's creator base in our sample skews heavily toward NSFW content (58\% by 2024), a category with distinct audience dynamics and willingness to pay. Twitter/X's historically social-graph-based content distribution, rather than pure algorithmic recommendation, is one plausible factor -- but so too are creator demographics, the platform's relatively weak native monetization infrastructure (which may select for creators with unusually loyal audiences), and the possibility that mid-tier creators on other platforms simply do not link to Patreon. We interpret Twitter/X's pattern as consistent with less aggressive algorithmic concentration, but cannot attribute it to algorithmic design alone. \rev{Accordingly, Twitter/X's favorable position in Figure~\ref{fig:scatter_median} may partly reflect NSFW audience willingness to pay rather than less concentrated algorithmic attention, limiting its usefulness as a pure benchmark for low-algorithmic-concentration outcomes.} In contrast, platforms with lower $\alpha$ show a ``rich-get-richer'' pattern far earlier on, where a small elite captures most of the income, leaving the middle-class of creators under-rewarded.

\begin{figure}[H]
	\begin{center}
      \caption{\centering {\large{Proportion of the earnings distribution following a power law pattern for each platform calculated on the pooled years.}}}
	\begin{minipage}{.79\textwidth}
		\includegraphics[width=\textwidth]{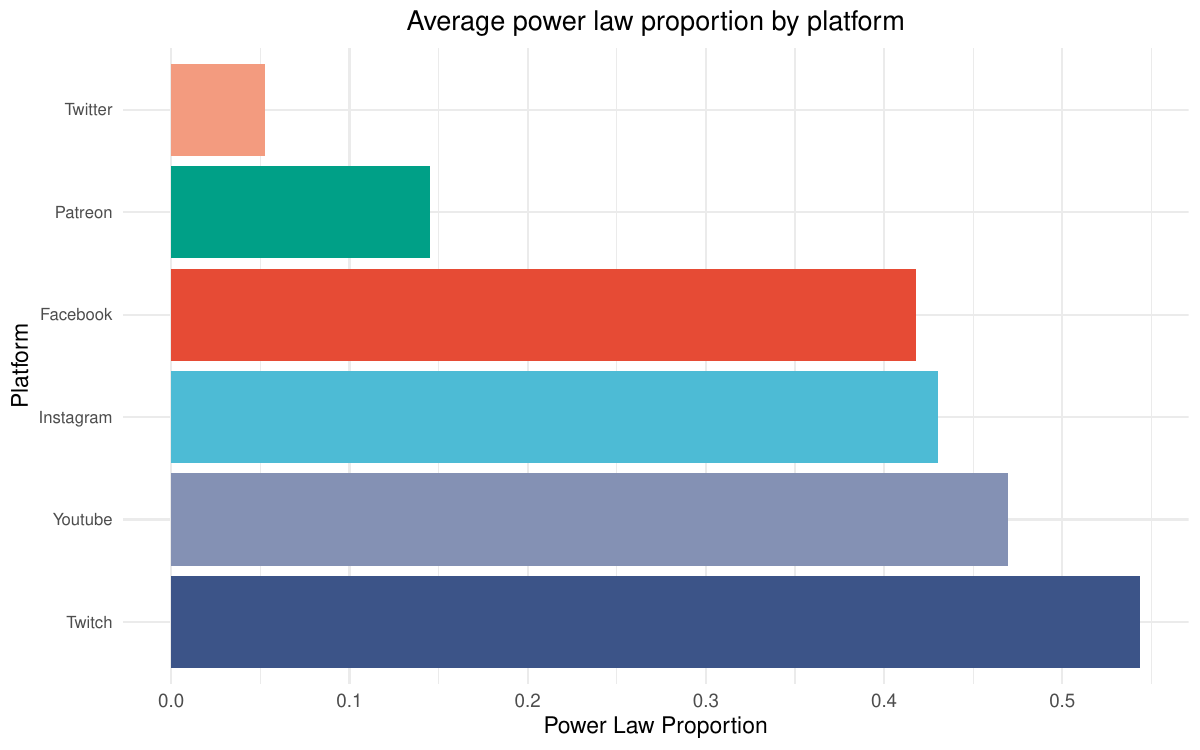}
	\end{minipage}

	\label{fig:powerprop}
	\end{center}
    \vspace{-3mm}
  \end{figure}

Figure~\ref{fig:powerprop} notes the proportion of observations for each platform (pooled across years) which follow a power law distribution. This allows us to assess the extent to which the power law distribution characterizes content creators' earnings. It shows that the proportion of earnings following a power law distribution varies significantly across platforms, from 5\% (Twitter/X) to almost 55\% (Twitch). Twitch and YouTube demonstrate the highest power law proportions (0.55 and ~0.47, respectively), suggesting that a large share of creator earnings is dominated by the power law distribution. This pattern is consistent with strong algorithmic recommendation systems and high reliance on viral content. Instagram and Facebook also show a relatively high and similar proportion (0.40-0.43). Patreon and Twitter/X show moderate to lower power law proportions, suggesting that only the very top of the income distribution aligns with the power law, as is often observed in standard income distributions \parencite{druagulescu2001exponential, tao2019exponential}.\\[-2mm]

\begin{figure}[H]
	\begin{center}
      \caption{\centering {\large{Average $\alpha$ by categories.}}}
      \vspace{-1mm}
	\begin{minipage}{.79\textwidth}
		\includegraphics[width=\textwidth]{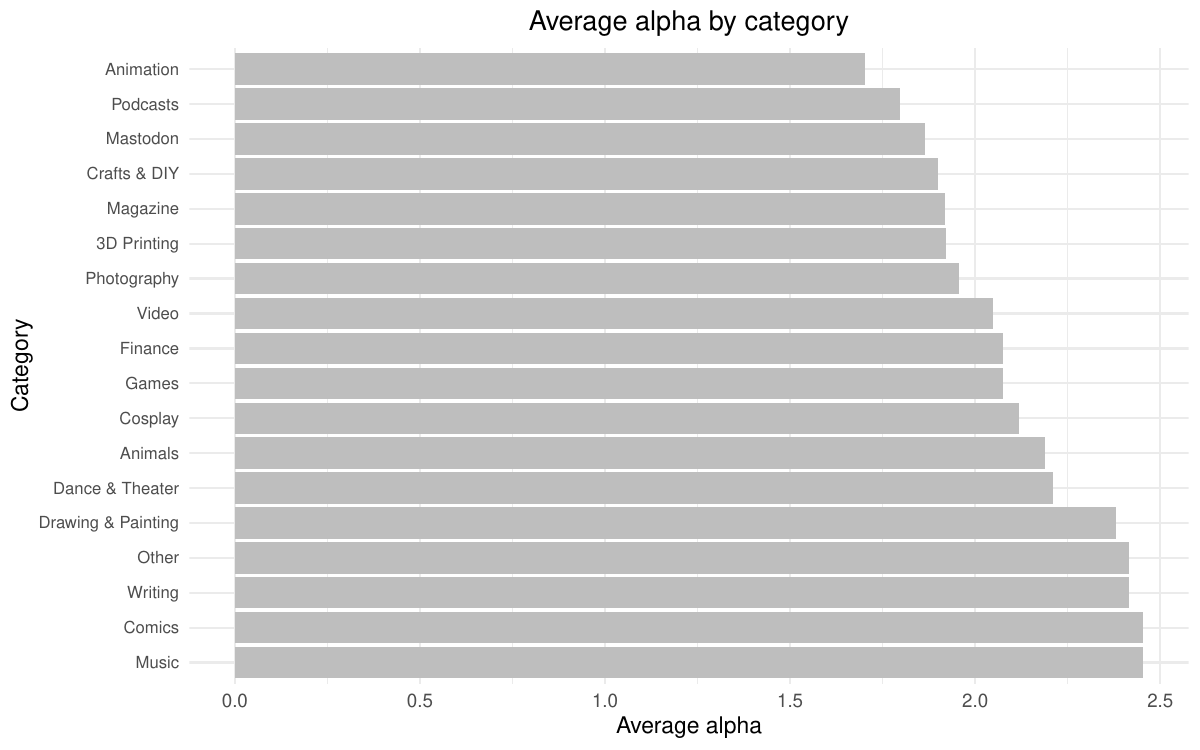}
	\end{minipage} 
  \caption*{\textit{Note}: The simple average represents the unweighted average across all categories for each year, while the weighted average is calculated based on the number of observations in each category.}
	\label{fig:cat}
	\end{center}
    \vspace{-4mm}
  \end{figure}

Lastly, Figure~\ref{fig:cat} illustrates the power law dynamics by Patreon defined category for each content creator, pooling across all social media platforms, including Patreon. Animation followed by Podcasts exhibit the lowest, i.e. most extreme, $\alpha$ values, indicating more pronounced earnings inequality compared to other categories. This may reflect how this content is disseminated and made viral, with a more extreme `winner-takes all' dynamic at play, possibly due to limited opportunities for audience discovery, the dominance of established creators, and its highly competitive consumption \& monetization structure. The Music, comics, and writing categories all show relatively high $\alpha$, suggesting more equitable earnings distributions. It may be harder to differentiate oneself in these categories as a content creator on Patreon, and may reflect a selection bias, with mid-tier content creators choosing to join Patreon in these categories but not the highest earners. These categories may provide better opportunities for mid-tier creators.\\[-2mm]

\section{Discussion: Market Structure, Algorithms, \& Policies}
\label{sec:Discussion}


Given the identification limits discussed in Section~\ref{sec:conversion_confound}, we position this paper primarily as a \emph{descriptive and methodological} contribution that documents cross-platform empirical regularities using Patreon as a novel lens. We treat algorithmic attention allocation as one plausible mechanism among several that could generate the observed patterns. Alternative explanations, including audience composition, creator sorting, and platform-specific conversion norms cannot be ruled out with cross-sectional data. Nevertheless, by detailing the power law dynamics governing digital creator economies, this paper contributes to broader discussions on fairness, sustainability, and innovation in networks \parencite{bianconi2001competition}. Our findings are consistent with a central role for algorithmic design in shaping earnings inequality and motivate future research that can more definitively identify causal mechanisms. 

\subsection{Platform market power and inequality in attention \& earnings}
The relationship between platforms' algorithmic choices over what content to prioritize and their market power has been theorized in the context of advertising and data privacy \parencite{srinivasan2019antitrust, strauss2024Amazon, o2024algorithmic}.\footnote{Market power reflects how competitive the social media market is—also known as the underlying `market structure'—defined by the number of firms in the market and shaped by cost structure and technology \parencite{carlton2015modern}.}

Similar dynamics appear to shape content creator earnings. In our data, Patreon-only and Twitter/X-only creators have more equal earnings distributions than those who monetize via more dominant platforms such as YouTube or Instagram. One interpretation is that less concentrated market structures are associated with less skewed algorithmic allocation of attention and earnings.

However, greater competition in the social media market has had ambiguous effects on the health of the ecosystem, including for content creators. Competition from TikTok appears to have locked the market into highly addictive, engagement-maximizing algorithms -- a race to the bottom. TikTok accelerated the shift away from the social graph towards purely algorithmic selection of content. While TikTok may promote more long-tail content, we lack evidence on whether this has translated into a more equal distribution of earnings within the platform, given the competitive pressures it faces \parencite{brennan2020attention}.

At the same time, competition has raised total payouts to creators as platforms bid for their participation \parencite{FT2023YouTubeShorts}. Yet this may increase within-platform inequality if top-tier creators are the main beneficiaries of competitive poaching. It also shows that creators do not always multi-home to the extent often assumed. Many focus on a single platform, as in our sample, so competition between platforms is only imperfectly transmitted to them, even if consumers do multi-home. In this context, greater interoperability -- via shared, open protocol infrastructures that allow creators to take followers with them across platforms -- may do more to reshape earnings dynamics than competition from platform entry alone.

So long as platforms retain strongly power-law–promoting algorithms, and monetize on that \textit{same} basis, earnings inequality is likely to persist. The more followers and views a creator has, the more they can earn \parencite{Naceva_2025_CreatorEP}. Gains to stardom are multiplicative rather than linear: outsized brand deals and advertising revenue shares accrue to a small group of superstar creators with enormous follower counts. It is notable that YouTube offers one of the most generous shares of ad sales, and hosts the greatest earnings inequality among Patreon creators who monetize there \parencite{FT2023YouTubeAdSharing}.

More competition alone is, therefore, unlikely to correct these unequalizing algorithms, especially if they dampen rational consumer choice and the usual benefits of competition on the user side. Interoperability mandates on the developer and creator side are likely a more promising route to raising the floor in creator earnings then, by making it easier for creators to move between competing platforms without leaving their audience behind.

\subsection{Alternative mechanisms}
\label{sec:alternative_mechanisms}

While we interpret our findings through the lens of algorithmic attention allocation, several alternative mechanisms could generate cross-platform differences in Patreon earnings inequality.

Platform audiences may differ systematically in their propensity to financially support creators through external platforms like Patreon. Instagram audiences, for example, may be less inclined toward direct patronage than Twitter/X audiences, independent of how attention is algorithmically distributed. Creators also self-select into platforms based on content type, audience fit, and monetization goals. The observed earnings distributions may thus partly reflect compositional differences in who uses each platform rather than (or in addition to) algorithmic effects.

The availability of substitute monetization channels further complicates interpretation. Ad revenue sharing, tipping, subscriptions, and brand deals vary across platforms. Creators on YouTube, for instance, may rely less on Patreon because YouTube's Partner Program offers substantial ad revenue, potentially creating selection effects in who monetizes via Patreon. Platforms also differ in how prominently creators can display external links, the friction users face when clicking out, and community norms around external monetization -- factors that can affect conversion rates independently of attention distribution. As documented in our data, platforms vary substantially in their treatment of adult content as well, leading to different creator compositions and potentially different earnings dynamics.

These mechanisms are not mutually exclusive with algorithmic explanations; indeed, algorithmic design may interact with each of them. However, our cross-sectional data cannot separately identify their contributions. Readers should interpret our findings as documenting \emph{patterns} in Patreon monetization outcomes rather than as definitive evidence for any single causal mechanism.

\paragraph{Creator sorting and compositional effects.}
An important alternative to the algorithmic explanation is that creators self-select into platforms in ways that generate the observed earnings patterns. If the most commercially ambitious creators concentrate on YouTube and Instagram while more niche creators gravitate toward Twitter/X or remain Patreon-only, the resulting earnings distributions would differ even under identical algorithms. Similarly, audience characteristics such as willingness to pay, cultural norms around patronage, demographic purchasing power vary across platforms and could independently shape Patreon earnings concentration. Our single-platform restriction mitigates some selection concerns (these are creators who have chosen \emph{one} external platform), but cannot eliminate them. The temporal trends we document are harder to attribute to compositional change alone, since they show within-platform shifts over three-year intervals, but we acknowledge that evolving creator and audience composition could contribute to these trends as well. Disentangling algorithmic effects from sorting and compositional effects would require either within-creator variation (e.g., the same creator switching platforms) or exogenous shocks to algorithmic design -- data we do not have but that would be valuable for future research.

\subsection{Policies to unlock more non-viral (longer-tail) content in algorithmic systems}
\rev{The policy recommendations in this section are conditional on algorithmic attention allocation being a first-order driver of the earnings concentration we document. As discussed in Sections~\ref{sec:conversion_confound} and~\ref{sec:alternative_mechanisms}, we cannot rule out alternative mechanisms, including: audience composition, creator sorting, and platform-specific conversion norms. If, however, algorithmic design is indeed a primary channel through which earnings become concentrated, then the following policy levers become relevant.}

A central challenge facing recommender systems on mature platforms is their potential to compound initial advantage (viewership), by promoting popular, existing material and focusing less on untested `fresh' content \parencite{covington2016deep, sutton2018reinforcement}. To the extent that this becomes baked into a platform's business model, with associated commercial incentives, it can become difficult to dislodge through market forces alone. 

A ``longer-tail'' algorithmic system \parencite{anderson2012long}, whereby a more diverse range of content is explored and promoted, invariably trades off some immediate profit from exploiting existing popular content in favor of greater content exploration \parencite{hu2023incentivizing, yao2024unveiling}. But maximizing for immediate user engagement \parencite{immorlica2024clickbait}, as a quarterly economic imperative facing the platform's shareholders, can lead to less new content being created and even less longer-term user engagement -- even if potentially at the expense of diminished immediate user satisfaction \parencite{facebook_notifications_2021, yao2024unveiling}. Several potential reforms are possible to grow healthier, more diverse and equitable, ecosystems of content promotion and consumption: 

Firstly, it remains to be seen if a new entrant into the market could improve what the market is currently optimizing for. Although the competitive dynamics of the market have seen a race to abandon the social graph and maximize time spent on platform through short-form videos, users also show a preference for longer-form content too as evidenced by the rise of lengthy podcasts. If the market needs a push in this direction then regulations that limit infinite scroll mechanisms would be one means of incentivizing algorithmic optimizations for longer-form content. But it is still unclear if this would benefit content creators and make for algorithmic recommendations that are more equalizing -- that tap into the long-tail more effectively.

Secondly, we recommend internal company measures that incentivize long-term orientated (algorithmic) A/B testing, to help demonstrate the business case for longer-term business optimizations at the algorithmic level \parencite{mazzucato_strauss_2024}. \textcite{hu2023incentivizing} recommends modifying the learning rates of recommendation algorithms to help balance the trade-off between exploration and exploitation, ensuring that high-quality content is recognized and then promoted quickly. Recent research suggests that incorporating mechanisms to surface underrepresented (``longer-tail'') content \parencite{anderson2012long} could enhance overall platform performance without necessarily sacrificing user satisfaction \parencite{houtti2024leveraging, yao2024unveiling}. 

Use of longer-tail discovery algorithms on social media to unlock and match a greater variety of non-viral content to users might also be incentivized through mandatory platform mechanisms that penalize the creation of low-quality content from dominant creators \parencite{hu2023incentivizing}. Facebook has recently chosen to do the opposite \parencite{techcrunch_2025}.

Thirdly, greater platform disclosure of the drivers of platforms' recommendation systems, in-line with the Digital Services Act (DSA, Article 27) of the European Union, may help with the adoptions of long-tail recommender systems \parencite{park2008long}. Platforms should be required to disclose key details about their recommendation algorithms to regulators, researchers, and the public. These disclosures should include the algorithmic weighting of past engagement metrics, such as views and likes, which often amplify the visibility of previously successful content. In addition, platforms should clarify the criteria by which new or less popular content is surfaced within the recommendation engine, ensuring greater accountability for algorithmic bias. In addition, mechanisms in place that punish low quality, click-bait content, should ideally be detailed. In the DSA mandated platform audits, the extent to which AI and ML recommendation systems favor certain types of creators, content genres, or user demographics should be evaluated. 

Platforms could allow users to customize their feeds, such as through ``content diversity sliders'' to adjust the balance between popular and niche content. YouTube, for example, sometimes tries to encourage this by asking if you want to explore unrelated topics to what you have been watching. And although Article 27 of the DSA requires platforms to offer users such ``non-profiling'' alternatives (e.g. chronological feeds), this is seen as an alternative to algorithmically curated ones. 

\section{Conclusion}\label{sec:conclusion}
Our analysis documents that Patreon earnings distributions across most major social media platforms closely follow power law dynamics, with an average exponent around $\alpha \approx 2.0$. This is a descriptive finding with potentially important implications: it places creator earnings closer to concentrated capital income than labor income, consistent with \rev{compounding ``rich-get-richer'' mechanisms. In the copying model \parencite{mitzenmacher2004brief, easley2010networks, chung2003generalizations}, $\alpha \approx 2$ corresponds to near-pure preferential attachment ($\gamma \to 0$) in a one-edge-per-arrival regime, where each impression or recommendation resolves into a single allocation of attention -- predominantly steered toward already-popular creators, with a small net exploratory component.} 

Importantly, platforms with stronger recommendation systems, such as YouTube ($\alpha = 1.8$) and Instagram ($\alpha = 1.84$), exhibit more significant earnings inequality and a weaker ``middle-class'' of creators (with lower mean, median, and interquartile earnings) -- a pattern consistent with amplified ``rich-get-richer'' dynamics. By contrast, Twitter/X, and to a lesser extent Patreon, with its direct creator-audience funding model, has a more equitable earnings distribution up until a certain point, with higher $\alpha$ values reflecting lower total earnings concentration, and a larger `middle-class' of creator earnings (higher mean, median, and interquartile earnings and the power law kicking-in much later in the earnings distribution).

Temporal trends show worsening (and converging) power law earnings dynamics of Patreon creators on Instagram, YouTube, Facebook, and Twitch, where decreasing $\alpha$ values suggest growing dominance of top Patreon creators. Instagram's $\alpha$ dropped from around 2.1 in 2021 to around 1.84 in 2024, a period that coincided with the platform's shift toward algorithmic recommendation. These trends are consistent with algorithmic systems increasingly favoring viral content over diverse creator engagement, though compositional shifts in the creator population could also contribute. 
Platforms with lower $\alpha$ values, such as YouTube and Instagram, face significant risks, including discouraging mid-tier creators and amplifying the spread of viral or harmful content. 

It is important to restate that our findings offer a partial and suggestive view of creator earnings dynamics. They document novel empirical patterns using a creative data source, and are consistent with the hypothesis that algorithmic attention allocation drives earnings concentration -- but they do not establish this causally. Alternative explanations, including creator self-selection, audience composition, and platform-specific monetization cultures, remain plausible and are not ruled out by our design. We view these findings as motivating future work with stronger identification, including studies exploiting within-platform algorithmic changes or cross-platform creator migration.

Finally, consideration of emerging AI technologies on algorithmic behavior and market structure are vital. AI technologies have for some time underpinned recommendation algorithms and advertising algorithms \parencite{covington2016deep, schafer_2023, gairola_2024}. If AI drives an explosion in content creation this risks reinforcing the dominance of existing platforms that intermediate and filter such content, as we come to rely on their algorithms for discovery even further. It may also lower barriers to entry for content creators, but it will not change online inequality and concentration in attention and earnings unless the underlying algorithmic mechanisms changes -- even if AI-generated content has the potential to foster more egalitarian ecosystems \parencite{zhou2024generative}. It's unclear at this stage how AI-native interfaces will change these dynamics.


\newpage                 
\begingroup
\setlength{\parindent}{0pt}  
\def\enotesize{\normalsize}  
\theendnotes
\endgroup

\pagebreak

\begin{appendices} \label{appendix}

\section{Robustness and Further Analysis}\label{app:robust}

\begin{table}[H]
\begin{center}
\caption{\centering {\large{Further statistics of Patreon monthly earnings by platform with NSFW}}} 
\begin{tabular}{llllll}
  \toprule
Platform & Year & Num\_Observations & Mean\_Earnings & Median\_Earnings & Sum\_Is\_Nsfw \\ 
  \midrule
Facebook & 2018 & 2,870 & 128 & 45 & 0.25 \\ 
  Facebook & 2021 & 3,068 & 170 & 52 & 0.18 \\ 
  Facebook & 2024 & 2,520 & 146 & 44 & 0.19 \\ 
  Instagram & 2018 & 414 & 185 & 60 & 0.46 \\ 
  Instagram & 2021 & 14,589 & 242 & 69 & 0.22 \\ 
  Instagram & 2024 & 20,876 & 216 & 53 & 0.27 \\ 
  Patreon & 2018 & 15,221 & 187 & 51 & 0.42 \\ 
  Patreon & 2021 & 30,768 & 287 & 63 & 0.36 \\ 
  Patreon & 2024 & 37,961 & 307 & 55 & 0.39 \\ 
  Twitch & 2018 & 54 & 136 & 41 & 0.43 \\ 
  Twitch & 2021 & 869 & 164 & 48 & 0.21 \\ 
  Twitch & 2024 & 870 & 236 & 42 & 0.21 \\ 
  Twitter & 2018 & 7,022 & 262 & 60 & 0.45 \\ 
  Twitter & 2021 & 16,576 & 382 & 77 & 0.46 \\ 
  Twitter & 2024 & 23,966 & 399 & 72 & 0.58 \\ 
  Youtube & 2018 & 2,143 & 199 & 44 & 0.17 \\ 
  Youtube & 2021 & 6,835 & 280 & 53 & 0.12 \\ 
  Youtube & 2024 & 14,457 & 243 & 43 & 0.12 \\ 
   \bottomrule
\end{tabular}

\label{tab:summary2}

\end{center}
\vspace{-4mm}
\caption*{\textit{Note}: Summary statistics by year and platform, with NSFW percentage.}
\end{table}

\begin{figure}[H]
	\begin{center}
      \caption{\centering {\large{Comparison of tail exponents from different estimation methods}}}
	\begin{minipage}{.9\textwidth}
		\includegraphics[width=\textwidth]{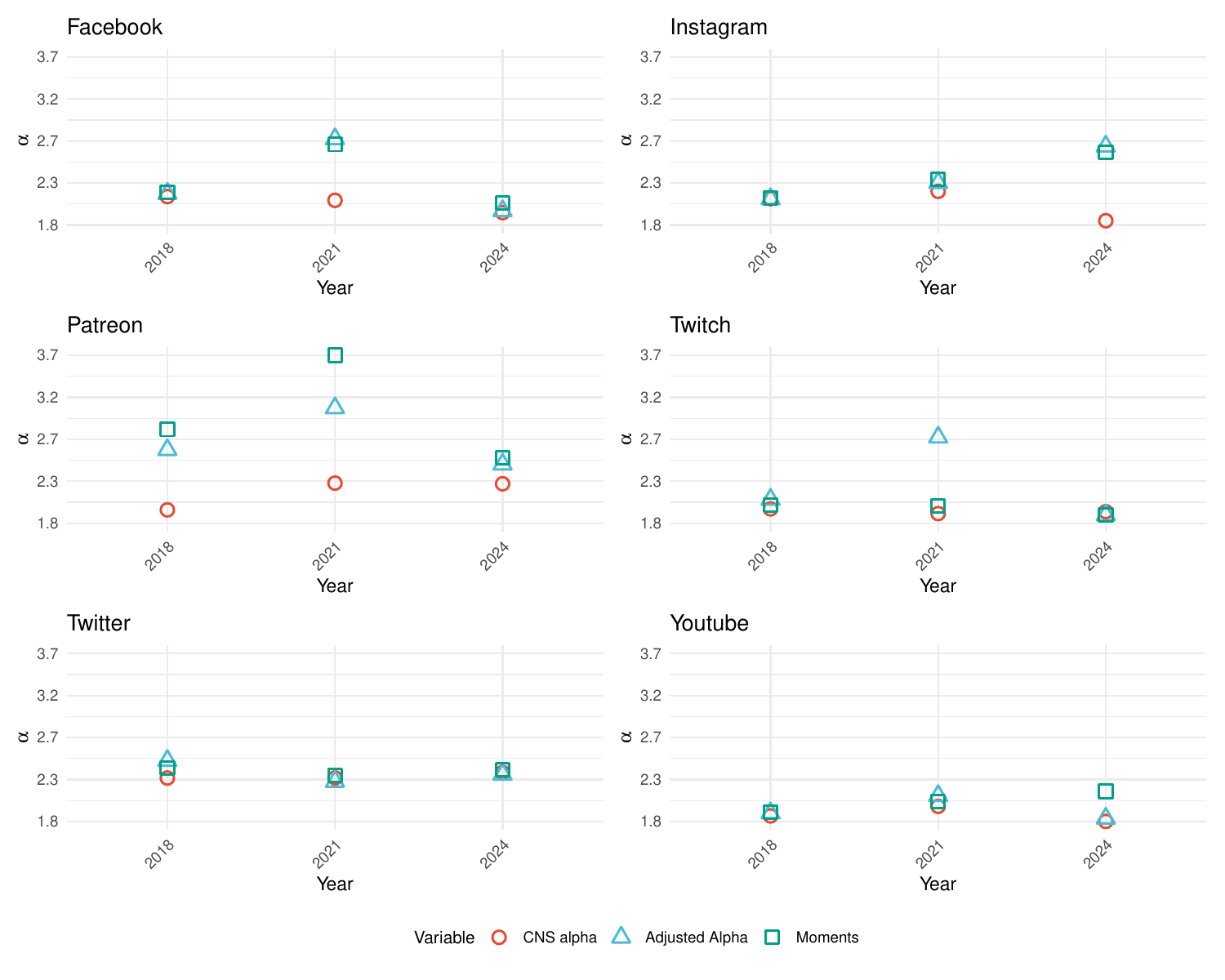}
	\end{minipage}
 \caption*{\textit{Note}: This graph compares tail exponents estimated using the Hill estimator, the Clauset-Newman-Shalizi (CNS) method \parencite{clauset2009power}, and the estimators described and implemented by \textcite{voitalov2019scale}. For the latter, the adjusted Hill estimator (Adj alpha) and the moments estimation (Moments) are used. The CNS method, as introduced by \textcite{clauset2009power}, relies on maximum likelihood estimation to fit the power-law model and uses the Kolmogorov-Smirnov (KS) statistic to assess goodness-of-fit. In contrast, \textcite{voitalov2019scale}'s implementation incorporates an automatic double bootstrapping procedure to determine the minimum value threshold.}
	\label{fig:robust}
	\end{center}
    \vspace{-2mm}
  \end{figure}
 Figure~\ref{fig:robust} compares tail exponents from three different estimation methods. With the exception of the earnings distribution for Patreon creators, the alpha values estimated by the three methods are generally consistent, confirming the robustness of the power-law pattern in the earnings distribution. For Patreon, discrepancies in the estimated exponents are particularly evident in the first two years (2018, 2021), where the tail decays more rapidly than the predicted line, as shown in Figure~\ref{fig:power_law_est}. Similarly, for the last year of the Instagram data, the CNS estimator deviates from the other two methods, again reflected in the faster decay of the tail relative to the predicted line. In nearly all other cases, the exponents estimated by the three methods are in strong agreement.

\section{Preferential Attachment Model}
 \label{app:power_law_der}
\subsection{Derivation of the relationship between $\alpha$ and $\gamma$}

We provide a heuristic derivation of the power law. Based on \textcite{mitzenmacher2004brief, easley2010networks, chung2003generalizations}, the derivation of the power law begins with an approximation to make the model tractable. We switch from a probabilistic to a deterministic framework, where we model the growth of links to a node as a continuous function of time, \(x_j(t)\), for node \(j\) created at time \(t=j\).

The growth rate of \(x_j(t)\) is described by the differential equation
\[
\frac{dx_j}{dt} = \frac{\gamma}{t} + (1-\gamma)\frac{x_j}{t},
\]
which combines two forces: a baseline flow of random links (scaled by \(\gamma\)) and a preferential term that is proportional to the current number of links (scaled by
\(1-\gamma\)). This is a first-order linear differential equation, which we can rewrite as
\[
\frac{dx_j}{dt} - \frac{1-\gamma}{t} x_j = \frac{\gamma}{t}.
\]
Using the integrating factor \(t^{-(1-\gamma)}\), we obtain
\[
\frac{d}{dt}\bigl[x_j(t)\, t^{-(1-\gamma)}\bigr] = \gamma\, t^{-(2-\gamma)}.
\]
Integrating with respect to \(t\) and imposing the initial condition \(x_j(j)=0\) (a node has no incoming links at birth) gives
\[
x_j(t) = \frac{\gamma}{1-\gamma}
\left[ \left(\frac{t}{j}\right)^{1-\gamma} - 1 \right].
\]
Thus, in this deterministic approximation, the expected number of links for node \(j\) increases as a power of the ratio between current time \(t\) and its creation time \(j\),
with exponent \(1-\gamma\).
 
The key insight from solving this model is that the fraction of nodes with a given number of links \(k\) at time \(t\) follows a power law distribution. Specifically, the probability that a node has \(k\) links decays as \(k^{-\alpha}\), where \(\alpha = 1 + \frac{1}{1-\gamma}\), a direct consequence of the rich-get-richer mechanism.

This derivation shows that the power law in network connectivity depends on the parameter \(\gamma\), which controls the balance between random and preferential linking. As \(\gamma\) decreases, emphasizing preferential attachment, the exponent \(\alpha\) approaches 2, indicating a stronger rich-get-richer effect with a more pronounced tail in the distribution.

\subsection{Derivation of the power law exponent in the original Barabási-Albert model}
\label{app:ba_model}

The preferential attachment model \parencite{barabasi1999emergence} is based on two mechanisms: growth, where a new node is added at each time step, and preferential attachment, where each new node connects to \( m \) existing nodes with a probability proportional to their degrees.

We denote the time as \( t \), with \( k_i(t) \) representing the degree of node \( i \) at time \( t \). The preferential attachment probability for a node \( i \) is given by

\begin{eqnarray}
\Pi(k_i) = \frac{k_i(t)}{\sum_{j} k_j(t)} = \frac{k_i(t)}{2mt},
\end{eqnarray}

where \( \sum_{j} k_j(t) = 2E(t) = 2mt \), since each new node brings \( m \) edges, and the total number of edges grows linearly with \( t \).

\subsection*{Degree evolution equation}
The expected increase in the degree of node \( i \) when a new node is added is

\begin{eqnarray}
\frac{dk_i(t)}{dt} = m \cdot \Pi(k_i) = m \cdot \frac{k_i(t)}{2mt} = \frac{k_i(t)}{2t}.
\end{eqnarray}

\subsection*{Solving the differential equation}
This is a separable differential equation. We can separate variables as follows:

\begin{eqnarray}
\frac{dk_i(t)}{k_i(t)} = \frac{dt}{2t}.
\end{eqnarray}

Integrating both sides:

\begin{eqnarray}
\int \frac{dk_i(t)}{k_i(t)} = \int \frac{dt}{2t},
\end{eqnarray}

which gives

\begin{eqnarray}
\ln k_i(t) = \frac{1}{2} \ln t + C.
\end{eqnarray}

Exponentiating both sides, we get

\begin{eqnarray}
k_i(t) = e^C \cdot t^{1/2}.
\end{eqnarray}

To determine the constant \( e^C \), note that at the time \( t_i \) when node \( i \) was added, it has degree \( k_i(t_i) = m \) (since it brings \( m \) edges upon introduction). Therefore,

\begin{eqnarray}
k_i(t_i) = m = e^C \cdot t_i^{1/2}.
\end{eqnarray}

Solving for \( e^C \), we find

\begin{eqnarray}
e^C = m \cdot t_i^{-1/2}.
\end{eqnarray}

Substituting back into the equation for \( k_i(t) \), we obtain the degree evolution equation:

\begin{eqnarray}
k_i(t) = m \cdot \left( \frac{t}{t_i} \right)^{1/2}.
\end{eqnarray}

\subsection*{Degree distribution}
From the degree evolution equation, solving for $t_i$, we get                 
                                                                                
  \begin{eqnarray}                                                              
  t_i = t \left( \frac{m}{k} \right)^2.                                         
  \end{eqnarray}                                                                
Since nodes are added uniformly over time, $P(t_i \le \tau)=\tau/t$. Hence,
\begin{eqnarray}
P(K \ge k)
&=& P\!\left(t_i \le t\left(\frac{m}{k}\right)^2\right)
 = \frac{t(m/k)^2}{t}
 = \left(\frac{m}{k}\right)^2,
\end{eqnarray}
which is the \emph{tail (complementary) CDF}. Equivalently, for $k\ge m$,
\begin{eqnarray}
P(K \le k)=1-\left(\frac{m}{k}\right)^2.
\end{eqnarray}

Differentiating yields
\begin{eqnarray}
p(k)=\frac{d}{dk}P(K\le k)=2m^2 k^{-3}\propto k^{-3},
\end{eqnarray}
so in the notation $p(k)\propto k^{-\alpha}$, the BA model gives $\alpha=3$.

\section{Missing Data and Imputation Validation} \label{sec:robust}

This appendix quantifies the extent of missing earnings data and evaluates the accuracy of the imputation method used in the analysis. Roughly one-third of creator-level earnings are missing. On Patreon in particular, creators can choose not to display their earnings publicly, and this choice is systematically related to their scale: higher-earning creators with more paid subscribers are disproportionately likely to hide their earnings. If we were to drop these observations, the right tail of the earnings distribution would be mechanically truncated, leading to a downward bias in both estimated inequality and the fitted power-law exponent.

To mitigate this, we impute missing earnings using an OLS model estimated on creators with observed earnings:

\[
\texttt{Earnings} \sim \texttt{Paid.Members} + \texttt{All.Members} + \texttt{Is.Nsfw} + \texttt{Year} + \texttt{Category},
\]

fit on complete cases (\texttt{na.omit}). The regressors capture the main observable drivers of earnings: the scale of the paying audience (\texttt{Paid.Members}), the broader follower base (\texttt{All.Members}), content type and NSFW status, and time effects. Since the probability of non-disclosure is strongly related to these observables (especially paid members), this specification approximates a setting in which earnings are missing at random conditional on the covariates. The fitted model is then used only to predict missing earnings; observed earnings remain unchanged.

Table~\ref{tab:missing} reports the sample size and missingness, alongside model performance. We show in-sample fit ($R^2$) and error magnitudes in levels. For interpretability, each error metric also reports its cross-validated counterpart in parentheses, computed via 10-fold out-of-sample prediction using the same specification. The model explains a substantial share of observed variation ($R^2\!=\!0.78$). Errors are moderate in absolute terms (RMSE $\approx\$547$, MAE $\approx\$81$), and the cross-validated errors are very similar, indicating limited overfitting. Given that our power-law estimates are driven by the broad shape of the upper tail rather than exact prediction of individual creators, this level of accuracy is sufficient for our purposes and preferable to discarding systematically missing high-earning observations.

\begin{table}[ht]
\centering
\caption{\large{Missing data and imputation diagnostics}}
\label{tab:missing}
\begingroup
\setlength{\tabcolsep}{5pt}
\renewcommand{\arraystretch}{1.1}
\resizebox{0.95\textwidth}{!}{%
\begin{tabular}{ccccccc}
  \toprule
N\_Tot & N\_Obs & N\_Miss & Miss (\%) & $R^2$ & RMSE (CV) & MAE (CV) \\
  \midrule
595,912 & 390,458 & 205,454 & 34.5\% & 0.778 & 547.17 (530.55) & 80.89 (81.17) \\
  \bottomrule
\end{tabular}%
}
\endgroup
\end{table}

\noindent\textit{Notes.} \texttt{N\_Tot}: total creators before imputation; \texttt{N\_Obs}: non-missing earnings; \texttt{N\_Miss}: missing earnings; \texttt{Miss (\%)}: share missing. Values in parentheses are 10-fold cross-validated (out-of-sample) metrics. RMSE = root mean squared error; MAE = mean absolute error;  absolute-error metrics are in USD.

\end{appendices}

\clearpage
\printbibliography
\end{document}